\documentclass[runningheads]{llncs}
\usepackage[T1]{fontenc}
\usepackage[english]{babel}
\usepackage{setspace}
\usepackage{lineno}
\usepackage{color}
\usepackage[T1]{fontenc}
\usepackage[section]{placeins}
\usepackage[pdftex]{graphicx}
\usepackage{floatflt}
\usepackage{float}
\usepackage{amsmath}
\usepackage{orcidlink}
\usepackage{mathtools}

\usepackage{array}
\newcolumntype{M}[1]{>{\centering\arraybackslash}m{#1}}
\newcolumntype{N}{@{}m{0pt}@{}}

\begin{document}

\title{Scientific Rigor and Human Warmth: Remembering Vladimir Sidorenko (1949–2025)}
\titlerunning{Remembering Vladimir Sidorenko}

\author{Christian Deppe\inst{1,6}\orcidlink{0000-0002-2265-4887} \and
Haider Al Kim\inst{3}\orcidlink{0000-0002-2429-1565} \and Jessica Bariffi \inst{2}\orcidlink{} \and Hannes Bartz \inst{5}\orcidlink{0000-0001-7767-1513}\and Minglai Cai\inst{1}\orcidlink{0000-0002-1327-2967}\and Pau Colomer\inst{2}\orcidlink{0000-0002-0126-4521}\and Gohar Kyureghyan\inst{4}\orcidlink{0000-0002-0126-4521}}
\authorrunning{Deppe, Al Kim, Bariffi, Bartz, Cai, Colomer, Kyurehgyan}
%
\institute{
Technical University of Braunschweig, Institute for Communications Technology, Braunschweig, Germany \and Technical University of Munich, TUM School of Computation, Information and Technology, Munich, Germany \and University of Kufa, Electronics and Communications Department (ECE), P.O Box 21, Kufa Street, Kufa, Iraq \and Universität Rostock, Institut für Mathematik,
18051 Rostock \and German Aerospace Center (DLR), Institute of Communications and Navigation, Oberpfaffenhofen-Wessling, Germany \and 6G-life, 6G research hub, Germany\\
\email{christian.deppe@tu-bs.de, haider.alkim@uokufa.edu.iq, jessica.bariffi@tum.de, Hannes.Bartz@dlr.de, minglai-cai@tu-bs.de, pau.colomers@tum.de, gohar.kyureghyan@uni-rostock.de}}

\maketitle

\begin{abstract}
During the Foundations of Future Communication Systems (FFCS) conference in Braunschweig, a dedicated memorial session was held in honor of Dr.\ Vladimir (``Volodya'') Sidorenko (1949--2025). The session, chaired by Minglai Cai, brought together colleagues, collaborators, and former students to commemorate his scientific achievements and his exceptional human qualities. This report summarizes the biographical tribute, the personal recollections shared by speakers, and the broader impact of Volodya’s work in coding theory, cryptography, telecommunications, and quantum error correction. Beyond his more than 150 publications and substantial technical contributions, the session highlighted his intellectual rigor, mentorship, humor, generosity, and lasting influence on the international research community.
\keywords{Vladimir Sidorenko \and Coding Theory \and Cryptography \and Quantum Error Correction \and Memorial Session}
\end{abstract}

\section{Introduction}

The FFCS conference included a memorial session dedicated to Dr.\ Vladimir Sidorenko, affectionately known to colleagues as ``Volodya.'' The session was chaired by Dr. Minglai Cai and began with a minute of silence in his honor. 

\begin{figure}[t]
    \centering
    \includegraphics[width=0.82\textwidth]{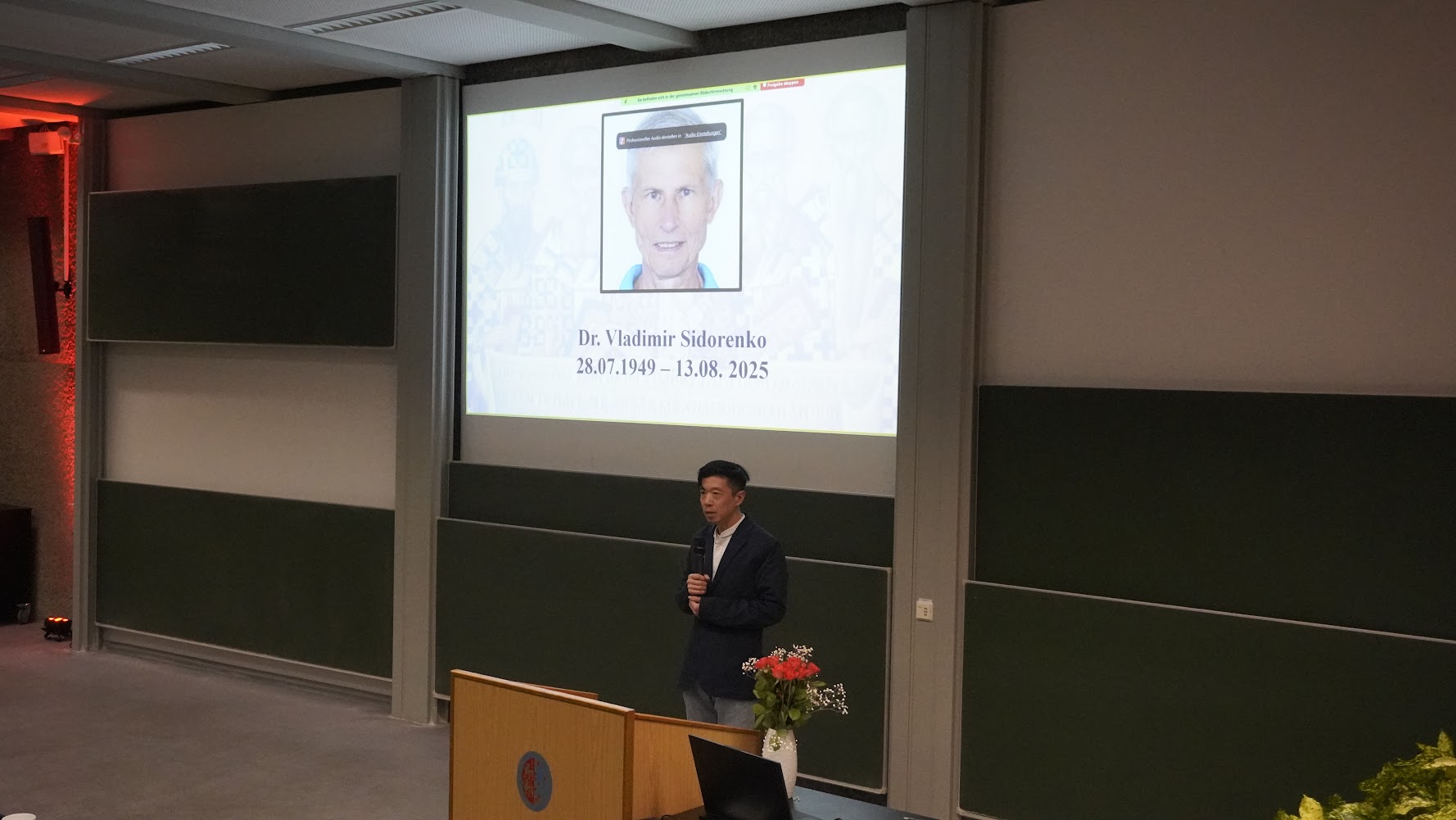}
    \caption{Minglai Cai opening the memorial session for Vladimir Sidorenko at FFCS.}
    \label{fig:minglai_intro}
\end{figure}

The gathering was not only an academic tribute, but also a deeply personal remembrance. Speakers reflected on Volodya’s scientific rigor, integrity, and intellectual curiosity, as well as his warmth, humor, and ability to bring people together. The session made clear that his legacy extends far beyond technical achievements: it lives on in the researchers he mentored, the collaborations he initiated, and the community spirit he cultivated.

This report documents the key elements of the session, beginning with a brief overview of his academic career and continuing with selected personal recollections shared during the event.

\section{Scientific Rigor, Curiosity, and Zest for Life: Reflections by Hannes Bartz}

\begin{figure}[t]
    \centering
    \includegraphics[width=0.82\textwidth]{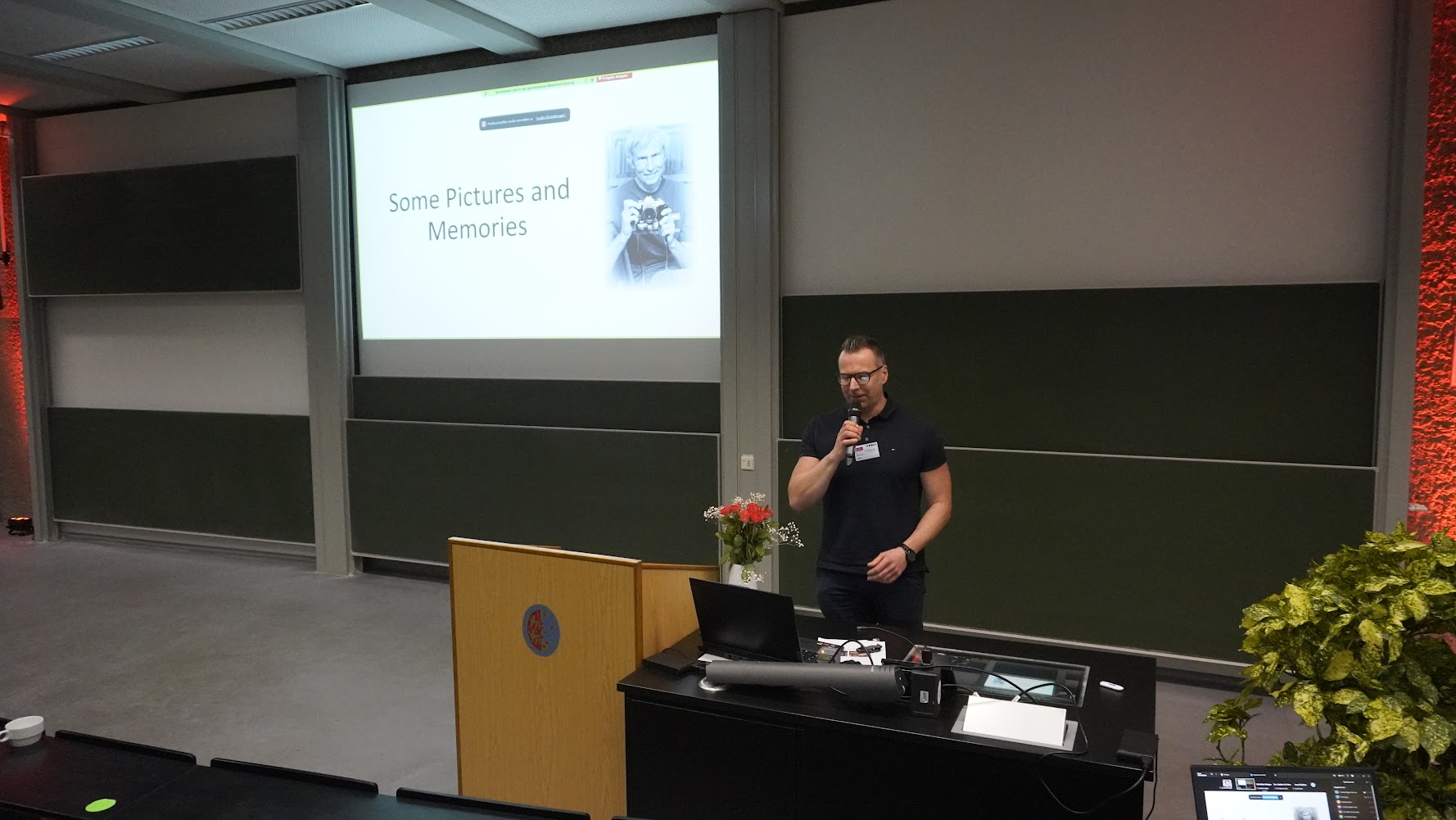}
    \caption{Hannes Bartz sharing his reflections during the memorial session.}
    \label{fig:hannes}
\end{figure}

In his contribution to the memorial session, Dr.\ Hannes Bartz (DLR) offered an extended and deeply personal reflection on his years of collaboration with Vladimir Sidorenko, particularly during the period when Volodya joined the Institute for Communications Engineering at TUM. At that time, Hannes was in the final phase of his doctoral studies, and their close scientific interaction left a lasting imprint on his academic development.

\subsection{Demand for Precision and Intellectual Honesty}

Hannes emphasized that one of Volodya’s most defining scientific traits was his uncompromising commitment to precision. He was deeply uncomfortable with vague arguments, heuristic reasoning, or claims that were not fully substantiated. In discussions, he repeatedly returned to definitions, assumptions, and logical structure. 

He often announced with a smile that he was about to ask a “nasty question.” These questions, however, were never destructive. They were constructive in the deepest sense: they exposed hidden assumptions, clarified ambiguous reasoning, and forced collaborators to confront weak points in their arguments. Hannes recalled how these repeated challenges—sometimes frustrating in the moment—ultimately led to clearer insights and stronger results.

In particular, when working on technically demanding problems, Volodya would revisit the same point multiple times from different angles. He would ask whether a statement held under slightly modified conditions, whether an argument was robust to edge cases, or whether a counterexample might exist. This habit of persistent questioning was not merely a stylistic feature; it reflected his conviction that good mathematics and good engineering require structural solidity.

For Hannes, this experience was formative. He learned not only how to refine proofs and constructions, but also how to approach research problems with intellectual humility and rigor. Volodya’s standard was simple but demanding: if a result is correct, it should withstand scrutiny from every direction.

\subsection{Passion for Conferences and Scientific Exchange}

Beyond the office and seminar room, Hannes highlighted Volodya’s enthusiasm for conferences. He was not a passive attendee; he was deeply engaged. He presented with energy and clarity, listened attentively to others, and frequently asked incisive questions during discussions.

Importantly, he was as interested in asking questions as in answering them. He appreciated precise results and elegant constructions, and he showed genuine excitement when encountering a new idea that was both conceptually clean and technically sound. His presence at conferences was noticeable—not because he dominated discussions, but because he animated them.

Hannes shared memories of joint conference trips, including meetings in St.\ Petersburg and various locations in Eastern Europe. These events combined technical programs with social interaction, and Volodya valued both dimensions equally. For him, scientific exchange was inseparable from human connection.

\subsection{Connecting People and Creating Community}

A recurring theme in Hannes’s recollections was Volodya’s gift for bringing people together. He enjoyed introducing colleagues to one another, especially when he sensed a potential intellectual match. He remembered names, research interests, and personal details, and he used this knowledge to create bridges within the community.

He also maintained longstanding connections with former colleagues and collaborators, such as Vladimir Lebedev and others from earlier stages of his career. These relationships were not purely professional; they were marked by genuine friendship.

According to Hannes, Volodya had a remarkable ability to create a positive atmosphere. His humor—often subtle and slightly ironic—helped ease tensions and encouraged open discussion. When collecting photographs for the memorial session, Hannes noticed a striking pattern: in almost every picture, Volodya was smiling. This visual continuity mirrored the emotional tone many associated with him.

\subsection{Childlike Curiosity and Optimism}

Among the anecdotes shared, one particularly vivid story involved a visit to Moscow during a conference hosted near the Olympic Village in Sochi. Volodya proudly showed Hannes his former desk at the Institute for Information Transmission Problems. While exploring the office, he rediscovered an old electrical device—a small toaster-like appliance.

With visible excitement, he decided to plug it in to demonstrate how it worked. Hannes, aware of the age and uncertain condition of the device, felt some apprehension. Yet Volodya reassured him with characteristic optimism: “Don’t worry, it will work.” Indeed, it did.

This episode, though seemingly minor, encapsulated an essential feature of his personality: a combination of technical confidence, curiosity, and an almost youthful delight in rediscovery. He approached both research and life with the same exploratory spirit.

\subsection{Sport, Music, and Vitality}

Hannes also described Volodya’s enthusiasm for sports and music. He was an active table tennis player, at times affiliated with a club in Ulm. During conference events that included recreational activities, he participated with energy and competitiveness.

Skiing was another passion. At conferences combining technical sessions with skiing programs, he was not merely a participant but often among the most energetic on the slopes. He also enjoyed ski touring and spoke proudly of learning to ski at a time when ski lifts were not yet available at his local hill. In those days, one had to hike uphill before each descent—a discipline that, as Hannes suggested, perhaps mirrored his perseverance in research: progress requires effort, repetition, and resilience.

Music formed yet another dimension of his life. He played the guitar and enjoyed singing. Informal musical moments at conferences revealed a different but entirely consistent aspect of his character: joy, expressiveness, and openness.

\subsection{Strength in Silence}

Hannes concluded his reflection by acknowledging the shock many felt upon learning of Volodya’s illness and passing. Characteristically, he had not widely shared details about his health situation. This discretion, Hannes suggested, was consistent with his personality: he preferred to focus on ideas, collaboration, and positive interaction rather than on personal hardship.

In looking back, Hannes emphasized that what remains are not only the papers and the theorems, but also the memories of laughter, discussion, and shared effort. The images of him smiling at conferences symbolize a legacy that is both intellectual and human.

Through his rigor, his questions, his humor, and his vitality, Vladimir Sidorenko shaped not only results in coding theory and cryptography, but also the researchers who continue to build upon them.

\section{Integrity, Mentorship, and Humility: Reflections by Haider Al Kim}

\begin{figure}[t]
    \centering
    \includegraphics[width=0.58\textwidth]{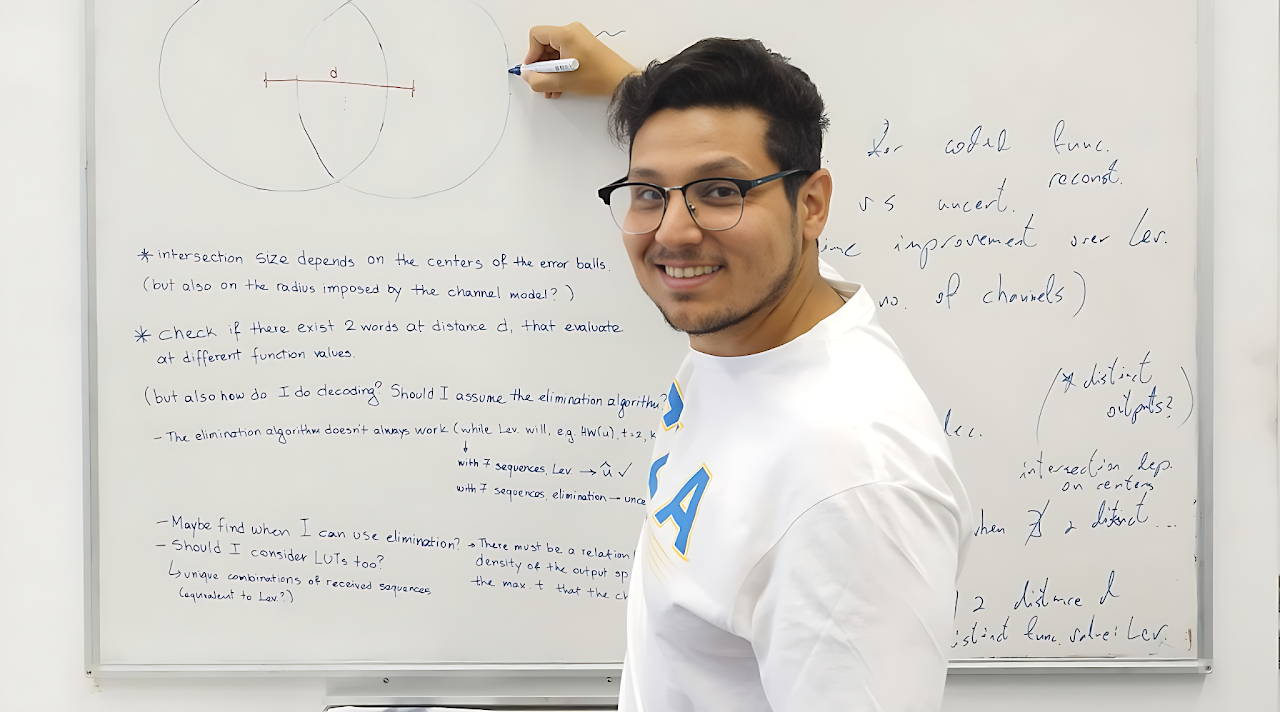}
    \caption{Haider Al Kim contributing to the memorial session via Zoom.}
    \label{fig:haider}
\end{figure}

Dr.\ Haider Al Kim, former doctoral researcher at the Technical University of Munich and now Assistant Professor at the University of Kufa, shared deeply personal reflections on his interactions with Vladimir Sidorenko. His contribution highlighted three essential aspects of Volodya’s character: uncompromising scientific integrity, genuine mentorship, and personal humility.

\subsection{A Mentor Beyond Formal Roles}

Although Volodya was neither Haider’s official doctoral supervisor—that role was held by Professor Dr. Antonia Wachter-Zeh—nor his official mentor, Dr. Sven Puchinger, he nevertheless played a central intellectual role in Haider’s development. Volodya frequently acted as a second mentor, engaging in detailed discussions across a broad range of topics in coding theory and communication engineering.

Haider emphasized that Volodya did not restrict himself to narrow specializations. He was curious about new directions, willing to explore unfamiliar ideas, and always ready to dedicate time to discussions. Even when his schedule was demanding, he would make space for technical conversations, carefully listening and challenging arguments where necessary.

What distinguished him as a mentor was not only his expertise, but his approach. He treated doctoral students not merely as junior collaborators, but as emerging colleagues. He encouraged independent thinking and intellectual ownership of ideas. At the same time, he expected rigor and clarity, setting high standards without imposing hierarchy.

\subsection{Authorship and Scientific Responsibility}

One of the central anecdotes shared by Haider concerned a joint research project on polyalphabetic codes. The work built upon Volodya’s earlier and well-known contributions in that area. His intellectual influence on the new paper was significant, and from a conventional academic perspective, co-authorship would have been natural.

However, during the development of the manuscript, Volodya expressed reservations about a specific modeling assumption that Haider initially wished to maintain. Although the assumption was not obviously incorrect, Volodya was not fully convinced of its necessity and conceptual soundness. Rather than compromise his standards, he made a striking decision: he preferred not to have his name on the paper as long as this assumption remained.

For Haider, this moment was transformative. As a doctoral student, authorship by a senior and highly respected researcher would have strengthened the visibility of the work. Yet Volodya’s priority was not visibility or recognition, but correctness. His stance was neither confrontational nor dismissive; it was principled.

Motivated by this position, Haider revisited the argument, re-examined the modeling choices, and revised the manuscript accordingly. The paper ultimately received strong and positive reviews. What impressed Haider most was Volodya’s reaction: he was genuinely happy---not because the paper was successful, but because the reasoning had been clarified and strengthened.

This episode encapsulated a core lesson: in scientific work, integrity precedes authorship. For Volodya, being listed as a co-author implied full intellectual endorsement. Anything less would have conflicted with his understanding of academic responsibility.

\subsection{Humility in Everyday Moments}

Beyond technical rigor, Haider shared a small but meaningful personal story illustrating Volodya’s humility and warmth. One day, Haider found a small ring and attempted to identify its owner. When he showed it to Volodya, he remarked that he believed it to be silver. Volodya responded with curiosity and amusement, openly admitting that he himself could not reliably distinguish silver.

Rather than asserting authority or dismissing the remark, he reacted with genuine interest and even admiration for Haider’s observation. The exchange, though minor in content, reflected an essential trait: Volodya never positioned himself above others in matters of everyday life. He approached interactions with openness and humor.

Despite efforts, the owner of the ring was never identified. Haider still wears the ring today as a quiet reminder of his mentor. The object has become a symbolic link to the conversations, the guidance, and the standards that shaped his doctoral journey.

\subsection{Encouragement and Intellectual Curiosity}

Haider also recalled Volodya’s excitement when exploring new ideas. Even in areas where he was already highly established, he remained curious and receptive. Discussions often extended beyond immediate research goals into broader questions of coding theory, combinatorics, and communication systems.

He did not impose solutions; instead, he asked questions that guided others toward discovering them. His mentoring style combined challenge and encouragement. He expected clarity, but he also celebrated progress---especially when younger researchers achieved conceptual breakthroughs.

In retrospect, Haider described him not only as a mentor, but as a model of how to conduct oneself in academia: principled without rigidity, critical without discouragement, and knowledgeable without arrogance.

\subsection{A Lasting Influence}

For Haider, the influence of Vladimir Sidorenko extends well beyond individual research projects. It shaped his understanding of what it means to be a scientist: to value correctness over credit, to engage deeply with ideas, and to treat colleagues and students with respect and humility.

The stories shared during the memorial session made clear that this influence was not isolated. Many in the community experienced similar guidance and encouragement. Through his integrity, intellectual generosity, and quiet humanity, Volodya left a legacy that continues in the work and values of those he mentored.

\section{Spontaneity, Generosity, and the Power of Informal Moments: Reflections by Pau Colomer}

\begin{figure}[t]
    \centering
    \includegraphics[width=0.82\textwidth]{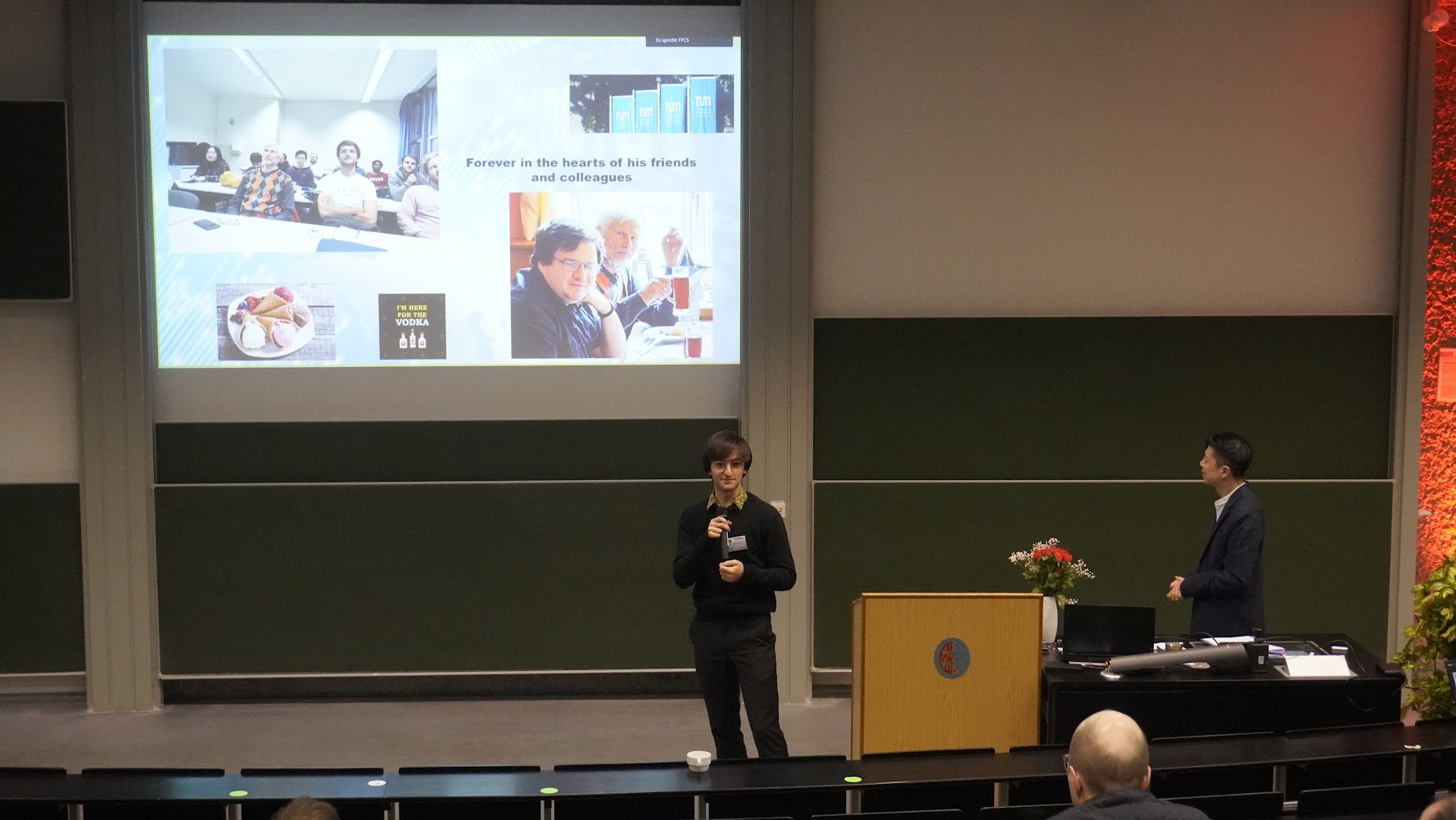}
    \caption{Pau Colomer speaking during the memorial session.}
    \label{fig:pau}
\end{figure}

Pau, doctoral researcher at the Technical University of Munich, offered a short but deeply meaningful recollection of his first encounter with Volodya. Although he emphasized that he had not known him for many years, the memory he shared illustrates how even brief interactions with Volodya could have a lasting impact.

\subsection{A First Week in Munich}

Pau arrived in Munich at the beginning of his doctoral studies, on the first of March, nearly three years prior to the memorial session. Having just relocated and still unfamiliar with most colleagues, he was invited by his mentor, Christian Deppe, to attend a workshop informally known as ``Coding and Skiing.'' The event combined a technical program in coding theory with extensive social interaction and recreational skiing.

For a newly arrived Ph.D.\ student, the prospect was both exciting and intimidating. It was his first opportunity to meet senior researchers, fellow doctoral students, and collaborators from different institutions. The atmosphere was intentionally informal, designed to foster discussion beyond strictly scheduled presentations.

\begin{figure}[t]
    \centering
    \includegraphics[width=0.82\textwidth]{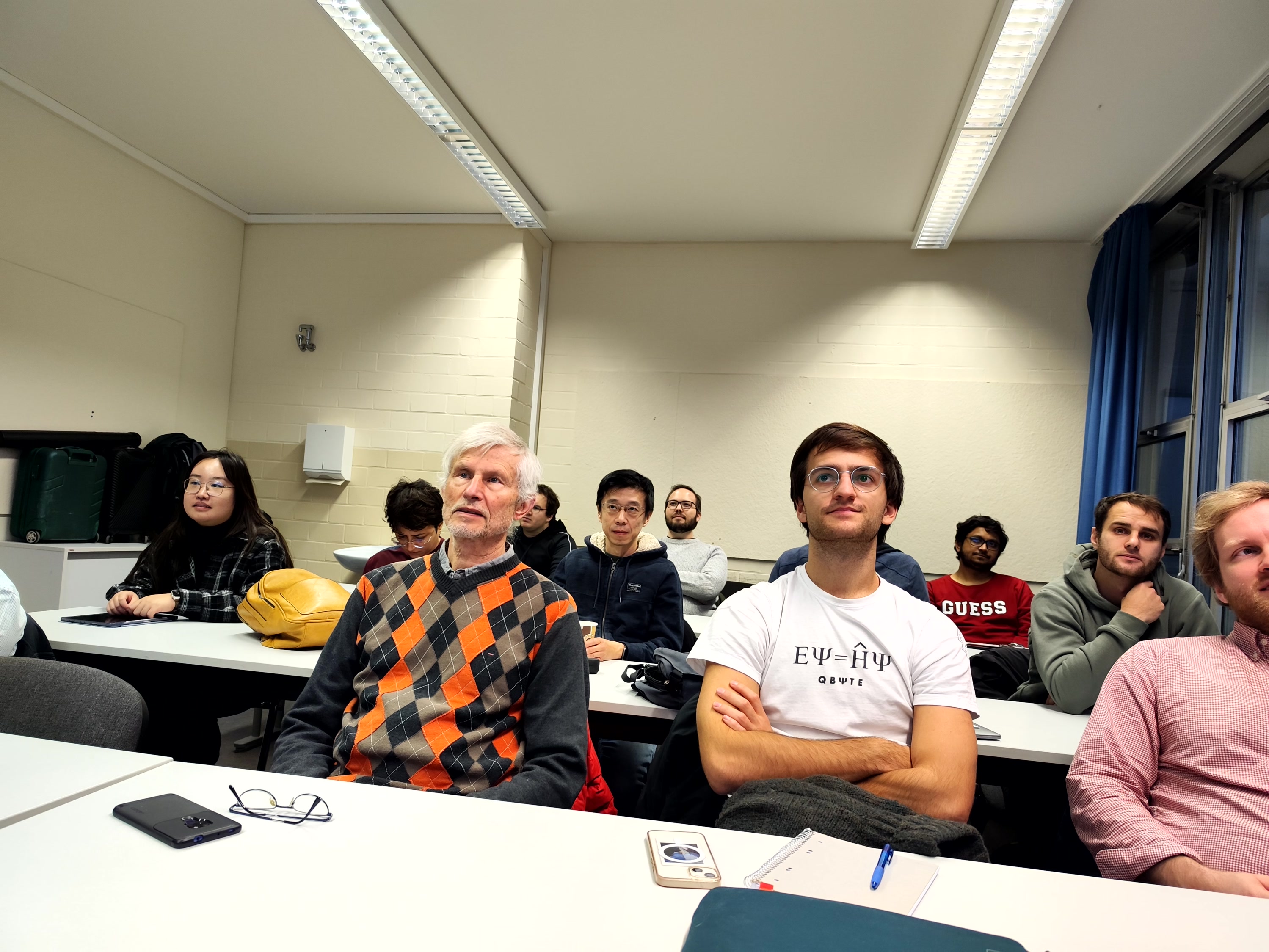}
    \caption{Pau Colomer with Vladimir Sidorenko in Bielefeld at the Post-Shannon Workshop in 2023.}
    \label{fig:pau_volodya}
\end{figure}

\subsection{The Dinner Table and a Bottle of Vodka}

Upon arrival in the afternoon, the participants gathered for dinner. Uncertain where to sit and unfamiliar with the social structure of the group, Pau simply followed his mentor and found himself seated at a table with several senior and highly respected researchers, including Volodya and Gerhard Kramer.

At that point, he did not yet know Volodya personally. What followed, however, became a defining memory. When dessert was served, Volodya spontaneously produced a bottle of vodka and began offering drinks to those at the table. Word of the free alcohol quickly spread among the students, and soon younger participants from across the room converged toward the table.

What might appear as a small gesture had a profound social effect. The bottle of vodka became a catalyst for conversation. Hierarchical boundaries dissolved. Senior researchers and first-year doctoral students shared stories, laughed together, and began forming connections that would later develop into professional collaborations and friendships.

Pau emphasized that some of the people he met that evening became among his closest colleagues in Munich. The informal gathering around that table helped him integrate into the research group far more quickly than any formal introduction could have done.

\subsection{Breaking Hierarchies, Building Community}

In retrospect, Pau recognized that this was not an isolated event. Others in the audience nodded in agreement, having experienced similar moments at conferences or institute gatherings. Bringing a bottle of vodka was not merely a habit; it was an instrument of community-building.

Volodya understood intuitively that scientific collaboration is sustained not only by shared technical interests, but also by trust, familiarity, and mutual respect. Informal moments—shared meals, spontaneous toasts, laughter—create the foundation upon which rigorous discussions can flourish.

Importantly, these gestures were never exclusive. He did not restrict his attention to established colleagues; rather, he intentionally drew students into the circle. For newcomers, such inclusion was invaluable. It conveyed the message that they were not merely observers, but members of the community.

\subsection{A Legacy of Human Connection}

Pau concluded his recollection with gratitude. Although he had interacted with Volodya only a limited number of times, the impact of that first encounter remained vivid. The memory of a spontaneous toast at a conference dinner symbolizes something larger: the generosity and openness that characterized Volodya’s approach to academic life.

The story also illustrates that legacy is not measured solely by publications, theorems, or citations. It is equally measured by the environments one helps create. In fostering moments of connection and breaking down barriers between generations of researchers, Volodya strengthened the very fabric of the scientific community.

For those who shared such evenings, the memory of his smile, his humor, and his instinct to gather people together remains inseparable from his scientific presence. Through these seemingly small acts, he shaped careers, friendships, and collaborations that continue to this day.

\section{Guidance, Openness, and Quiet Dedication: Reflections Shared by Jessica Bariffi}

\begin{figure}[t]
    \centering
    \includegraphics[width=0.82\textwidth]{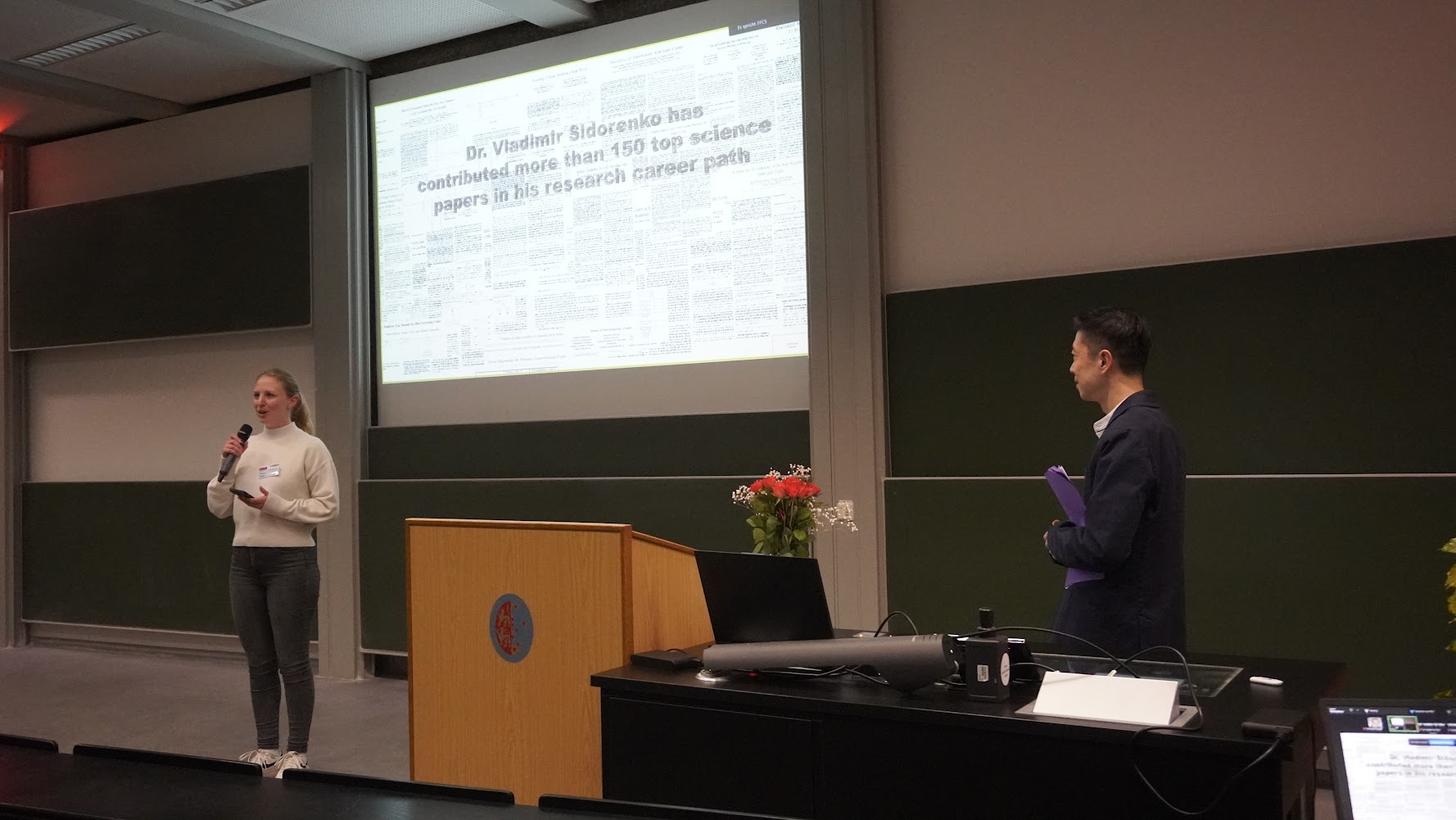}
    \caption{Jessica Bariffi sharing reflections during the memorial session.}
    \label{fig:jessica}
\end{figure}

Dr. Jessica Bariffi, postdoctoral researcher at the Technical University of Munich, spoke on behalf of her colleague Evagoras Stylianou who was unable to attend the memorial session. Although she explained that she had not worked closely with Volodya herself, the story she conveyed---together with her own impressions---offered a vivid portrait of his mentorship, intellectual openness, and human warmth.

\subsection{Supporting the First Steps of a Ph.D.}

The recollection began with the early phase of Evagoras a doctoral project under the supervision of Christian Deppe. The research topic involved quantum error correction, a technically demanding and conceptually challenging area, especially for someone without prior background in quantum information theory. The initial stage of the Ph.D.\ was therefore marked by uncertainty and considerable pressure to identify a viable research direction.

At this crucial moment, Volodya approached the young researcher and suggested that they learn the subject together. Despite being widely recognized for his expertise in classical coding theory, he did not hesitate to engage deeply with quantum topics. Rather than positioning himself as an authority delivering ready-made solutions, he offered partnership: they would explore the theory jointly, identify connections to classical coding, and search for meaningful applications of his expertise.

\begin{figure}[t]
    \centering
    \includegraphics[width=0.82\textwidth]{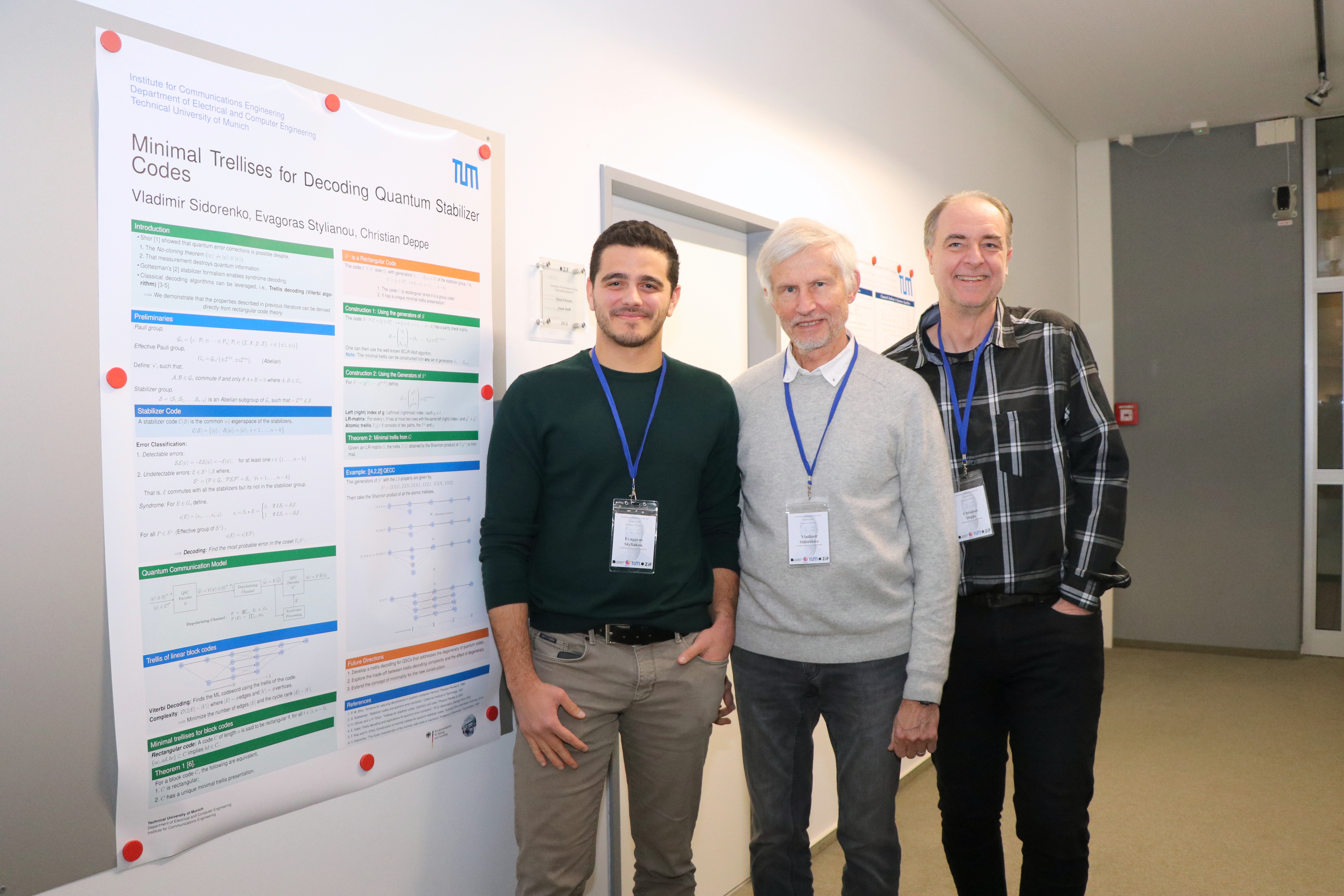}
    \caption{Evagoras Stylianou with Vladimir Sidorenko and Christian Deppe presenting their joint work \cite{Stylianou2025MinimalTrellisITRF}.}
    \label{fig:evagoras_volodya}
\end{figure}

This gesture was emblematic. It demonstrated intellectual courage, the willingness to step into a new domain, and pedagogical generosity. For a beginning Ph.D.\ student, the message was clear: uncertainty is not a weakness; it is the starting point of research.

\subsection{From Classical to Quantum Coding}

The collaboration eventually focused on the problem of degenerate decoding of quantum stabilizer codes using trellis-based methods. The project aimed to transfer classical coding-theoretic insights into the quantum setting, carefully adapting concepts and tools.

Although the work was only partially completed, and plans for further extensions remained unrealized, the intellectual process itself was formative. Volodya’s approach was characterized by patient explanation, persistent questioning, and an effort to build conceptual bridges between classical and quantum perspectives.

He did not merely provide technical input; he helped structure the problem. He asked which assumptions were essential, which analogies were legitimate, and where classical intuition might fail in the quantum regime. Through this method, he modeled a way of thinking that was analytical yet creative.

\subsection{Human Warmth in Professional Contexts}

Jessica also conveyed memories of Evagoras that went beyond research discussions. During online meetings, especially in the early stages of the collaboration, glimpses of Volodya’s family life were visible. His son would occasionally appear in the background, and Volodya would seamlessly combine professional discussion with parental attention. These moments, though informal, underscored his authenticity and grounded presence.

At institute events, he made deliberate efforts to attend, even when travel was inconvenient. He sometimes joked that he came for the free food, but it was evident that he valued conversation and connection. Social gatherings were not peripheral to him; they were integral to academic life.

One particularly memorable anecdote involved a farewell event at the institute. Due to a train strike, he was forced to leave shortly after arriving. Before departing, however, he approached a colleague with a serious expression and entrusted her with a “mission”: a bottle of his favorite Russian vodka was stored in the refrigerator, and she was to ensure that everyone had a shot in his absence. Moreover, she was instructed to finish the entire bottle, no excuses permitted.

The episode, humorous in tone, reflected both his sense of playfulness and his desire to remain part of the communal moment, even when physically absent.

\subsection{Presence Without Self-Promotion}

A recurring theme in Jessica’s contribution of Evagoras memories was that Volodya never sought attention for himself. His guidance was offered quietly. He did not dominate discussions or emphasize his seniority. Instead, he encouraged others to develop confidence in their own ideas.

He combined high scientific standards with patience. When confronted with incomplete understanding or conceptual confusion, he responded not with criticism, but with curiosity. He asked clarifying questions, proposed alternative viewpoints, and gently guided the conversation toward clarity.

Such an approach left a lasting impression on young researchers. It demonstrated that intellectual rigor and kindness are not mutually exclusive; they can and should coexist.

\subsection{A Memory of Joy and Dedication}

Jessica concluded by emphasizing that what many colleagues remember most vividly is not a single theorem or paper, but the atmosphere Volodya created: relaxed, humorous, and intellectually vibrant. He attended conferences enthusiastically, engaged in discussions across subfields, and welcomed collaboration.

Through mentorship, shared exploration of new topics, and small gestures of inclusion, he helped shape the academic journeys of numerous students and postdoctoral researchers. Even projects that remained unfinished contributed to intellectual growth and confidence.

In reflecting on these stories, it becomes clear that Volodya’s legacy is woven into the development of those he supported. His dedication to research, combined with authenticity and warmth, continues to influence the community he helped build.

\section{Humanity, Encouragement, and Lasting Kindness: Reflections by Gohar Kyureghyan}

\begin{figure}[t]
    \centering
    \includegraphics[width=0.82\textwidth]{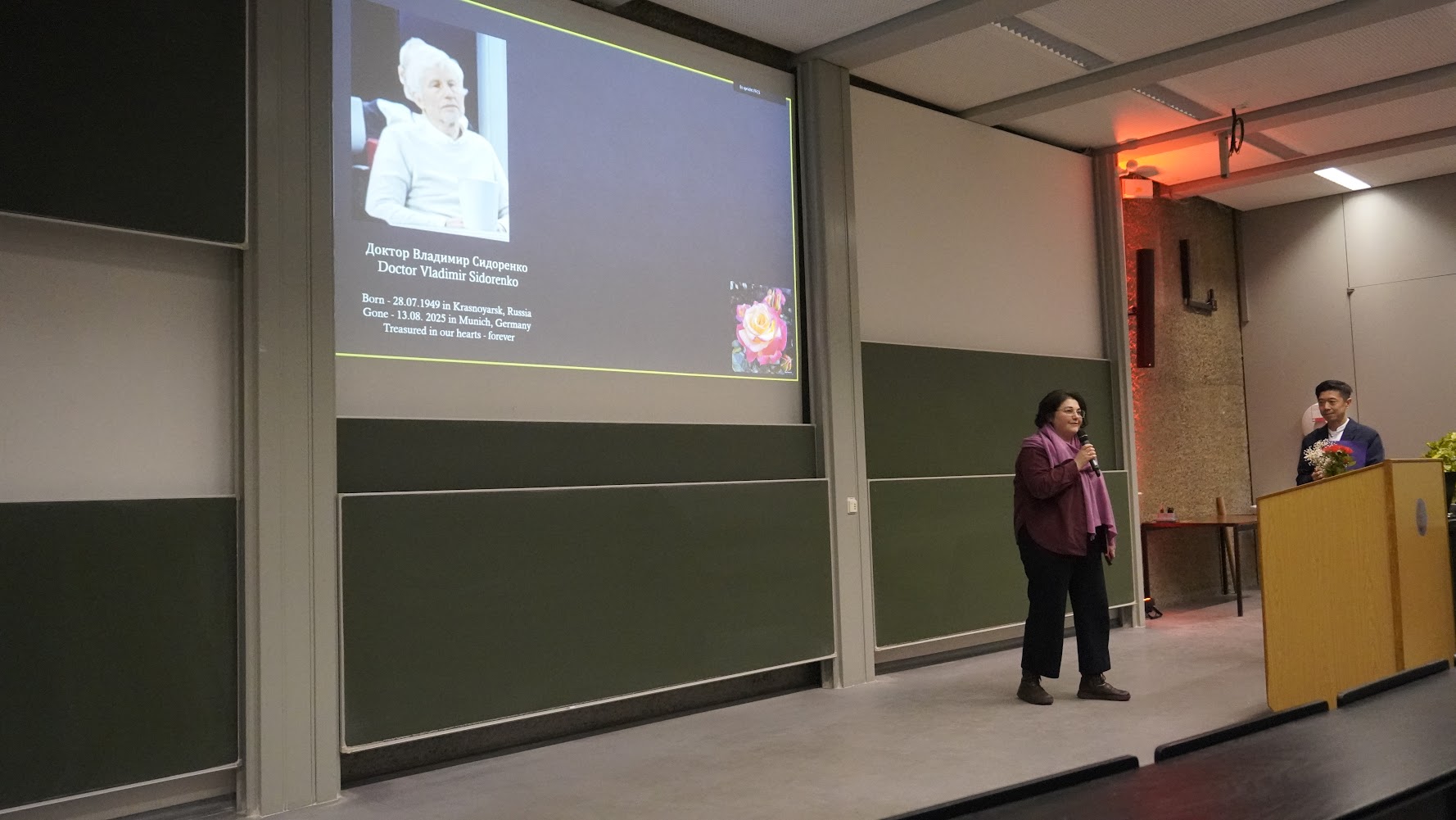}
    \caption{Gohar Kyureghyan during her contribution to the memorial session.}
    \label{fig:gohar}
\end{figure}

Professor Gohar Kyureghyan (University of Rostock) contributed a deeply personal and moving reflection that highlighted both Vladimir Sidorenko’s intellectual stature and his human warmth. Although she did not collaborate with him in all areas of his work, his presence left a lasting impression on her academic life and on the broader community.

\subsection{The Colleague Who Could Review Everything}

One of her earliest memories of Volodya arose during the organization of a scientific conference. While discussing potential members of the program committee, a senior researcher---whose judgment she greatly respected---remarked that Volodya absolutely had to be included because ``he can review all the difficult papers.''

This statement, brief yet telling, encapsulated a defining aspect of his scientific reputation. He possessed exceptional breadth and depth. Colleagues trusted him not only for his technical competence, but for his ability to penetrate complex arguments and evaluate them with fairness and clarity.

Professor Kyureghyan noted that Volodya was known for asking what he himself sometimes called ``nasty questions.'' Yet these questions were never destructive. On the contrary, they were constructive in the strongest sense. They identified hidden assumptions, clarified ambiguous reasoning, and strengthened arguments. Such questioning, she emphasized, is essential for scientific progress. A research community grows when its members challenge one another respectfully and rigorously. Volodya fulfilled this role with intellectual honesty and deep respect for mathematics.

\subsection{A Cup of Tea}

While his scientific rigor was widely recognized, Professor Kyureghyan chose to center her remembrance on a personal episode that revealed his kindness.

At one conference, she had been invited to give a talk, one of her first invited lectures. Nervous and determined to perform well, she was at the same time suffering from a severe cold. Shortly after beginning her presentation, she was overcome by persistent coughing. Despite having prepared water to ease the situation, the coughing worsened. Approximately fifteen minutes into the talk, she noticed Volodya quietly leaving the room. In that moment, she felt discouraged and interpreted his departure as a sign that the situation had become unbearable.

However, a few minutes later he returned carrying a cup of warm tea.

He gently explained that drinking cold water could aggravate the cough and that warm tea would help calm it. Thanks to this small but thoughtful gesture, she was able to continue and complete her talk. At the end, he reassured her that it had been a good presentation.

For Professor Kyureghyan, this moment became emblematic. Beyond being an outstanding mathematician, Volodya was attentive, empathetic, and quietly supportive. His kindness did not manifest in grand gestures, but in small acts at precisely the right time.

\subsection{Humanity in the Scientific Community}

In her closing remarks, Professor Kyureghyan reflected on the importance of humanity within academic life. Excellence in mathematics and engineering is indispensable, but communities are sustained by more than technical achievements. They require generosity, encouragement, and mutual care.

When she thinks of Volodya, she remembers not only his sharp analytical mind, but also his smile and his kind eyes. These images, she said, are inseparable from his scientific presence.

Her contribution reminded the audience that his legacy is twofold: intellectual rigor and human warmth. The standards he upheld in research, combined with the empathy he demonstrated in everyday interactions, continue to shape the community he helped build.

\begin{figure}[t]
    \centering
    \includegraphics[width=0.82\textwidth]{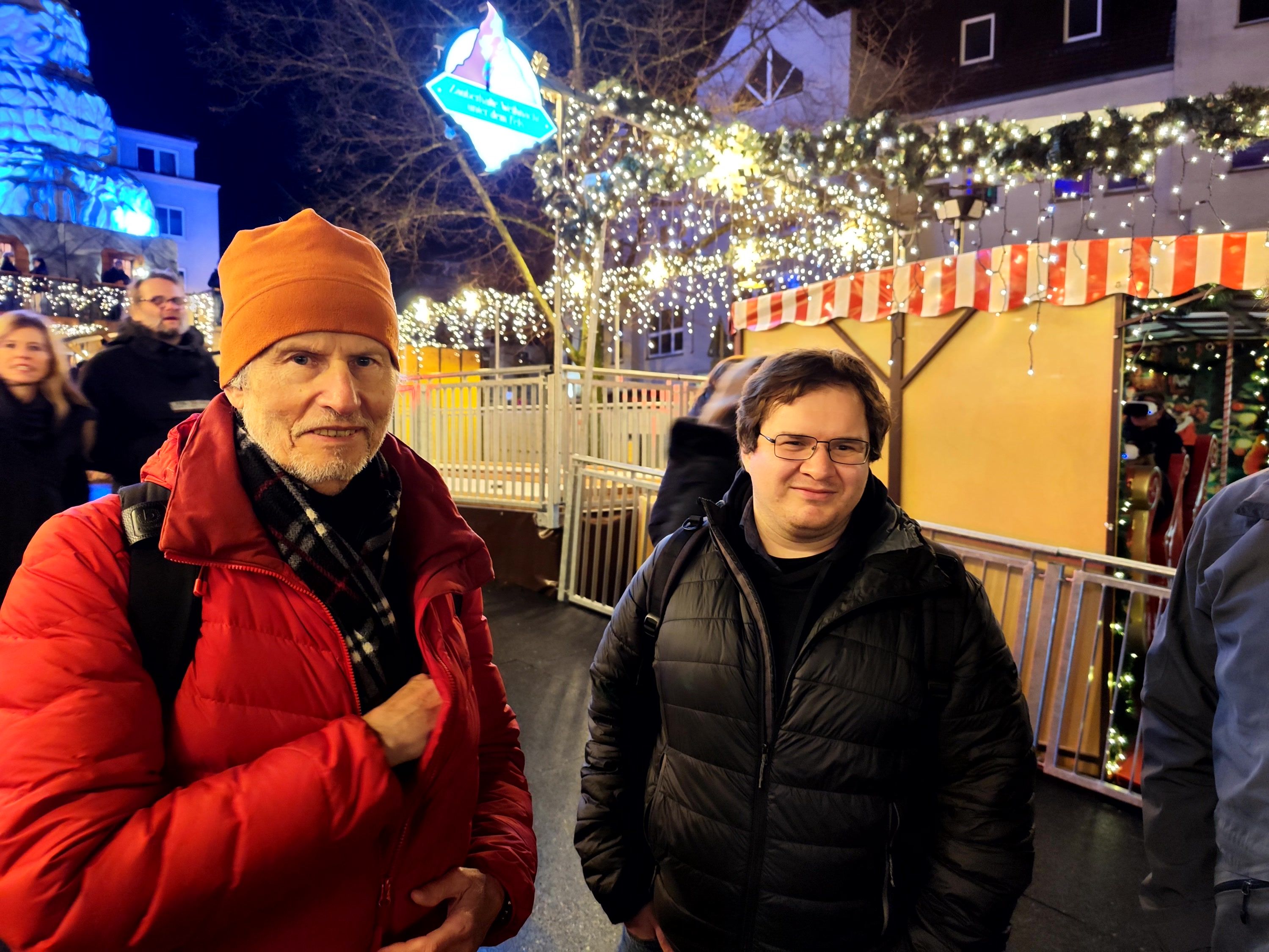}
    \caption{Vladimir Sidorenko with his colleague Ilya Vorobyev.}
    \label{fig:ilya_volodya}
\end{figure}

\section{Conclusion}

The memorial session made clear that Vladimir Sidorenko’s legacy is both scientific and human. His contributions to coding theory, cryptography, telecommunications, and quantum error correction remain influential. Equally enduring, however, are the standards of rigor, integrity, and collegiality that he embodied.

A special issue dedicated to algebraic and combinatorial methods in coding and cryptography has been initiated in his memory, reflecting the broad impact of his work. 

The session concluded with a shared understanding: while the community has lost an exceptional researcher and colleague, his influence continues through his publications, his students, and the many researchers who carry forward his commitment to precise thinking and generous collaboration.

\bigskip

We end this report with the publications of Volodya.

\nocite{*}

\bibliographystyle{splncs04}
\bibliography{references}

\end{document}